\documentstyle[aps,preprint]{revtex}

\title{Comparison of Fermion SU(3) and Boson SU(3) models for scissors
mode excitations}

\author{Y. Y. Sharon\cite{byline}, L. Zamick}
\address{Department of Physics, Rutgers University,
Piscataway, New Jersey 08855}
\author{M. S. Fayache}
\address{Department de Physique, Facult\'{e} des Sciences de Tunis,
Tunis 1060, Tunisia}
\author{G. Rosensteel}
\address{Department of Physics, Tulane University,
New Orleans, Louisiana 70118}

\begin{document}

\maketitle

\begin{abstract}
For a $Q \cdot Q$ interaction the energy weighted sum rule for
isovector orbital magnetic dipole transitions is proportional to the
difference $\sum B(E2, {\rm isoscalar}) - \sum B(E2, {\rm isovector})$, 
not just to $\sum B(E2, {\rm physical})$.  This fact is important in 
ensuring that one gets the 
correct limit as one goes to nuclei, some of which are far from
stability, for which one shell (neutron or proton) is closed.
In $0p$ shell calculations for the even-even
Be isotopes it is shown that the Fermion
SU(3) model and Boson SU(3) model give different results for
the energy weighted scissors mode strengths.
\end{abstract}

%\narrowtext

\section{Energy Weighted sum rule for scissors mode excitations in the
fermion SU(3) model}

Using the interaction $-\chi Q \cdot Q$, Zheng and
Zamick~\cite{Zamick-L:1991prc,Zamick-L:1992prc}
obtained a sum rule which relates the scissors mode excitation rate
(i.e. the isovector orbital magnetic dipole excitation rate) to the
electric quadrupole excitation rate.  The isovector orbital magnetic
dipole operator is $(\vec{L}_{\pi}-\vec{L}_{\nu})/2$ (the isoscalar one
is half the total orbital angular momentum $\vec{L}/2 =
(\vec{L}_{\pi}+\vec{L}_{\nu})/2$).  In more detail, the sum rule reads
\begin{eqnarray}
\lefteqn{\sum_{n} (E_{n} - E_{0}) B(M1)_{o} = } && \nonumber \\
 && \;\;\; {9 \chi \over 16 \pi} \sum_{i} \left\{\left[
   B(E2, 0_{1} \rightarrow 2_{i})_{IS} -
   B(E2, 0_{1} \rightarrow 2_{i})_{IV} \right]\right\}
  \label{eq:ewsr}
\end{eqnarray}
where $B(M1)_{o}$ is the value for the {\em isovector} orbital M1
operator ($g_{l\pi} = 0.5, g_{l \nu} = -0.5, g_{s \pi} = 0, g_{s \nu}
= 0$) and the operator for the E2 transitions is 
$\sum_{{\rm protons}} e_{p} r^{2} Y_{2} + \sum_{{\rm neutrons}} e_{n}
r^{2} Y_{2}$ with $e_{p} = 1$, $e_{n} = 1$ for the isoscalar
transition (IS) and $e_{p} = 1, e_{n} = -1$ for the isovector
transition (IV).  The above result also holds if we add a pairing
interaction between like particles, i.e between two neutrons and two
protons.

The above work was motivated by the realization from many sources that
there should be a relation between the scissors mode excitation rate
and nuclear collectivity.  Indeed, the initial picture by Palumbo and
LoIudice~\cite{LoIudice-N:1978prl} was of an excitation in a deformed
nucleus in which 
the symmetry axis of the neutrons vibrated against that of the protons.
In 1990-91 contributions by the Darmstadt
group~\cite{Ziegler-W:1990prl,Ranacharyulu-C:1991prc}, it was noted
that the Sm isotopes, which undergo large changes in deformation as a
function of mass number, the
$B(M1)_{{\rm scissors}}$, was proportional to $B(E2, 0_{1} \rightarrow
2_{1})$.
The $B(E2)$ in turn is proportional to the square of the nuclear
deformation $\delta^{2}$.

The above energy weighted sum rule of Zheng and
Zamick~\cite{Zamick-L:1992prc} was an attempt
to obtain such a relationship microscopically using fermions rather
than interacting bosons.  To a large extent they succeeded, but there
were some differences relative
to~\cite{Ziegler-W:1990prl,Ranacharyulu-C:1991prc}.
Rather than being proportional to $B(E2, 0_{1} \rightarrow 2_{1})$, 
the proportionality factor was 
the {\em difference} in the summed isoscalar and summer isovector $B(E2)$'s.
Now one generally expects the
isoscalar $B(E2)$, especially to the first $2^{+}$ state,
to be the most collective and much larger than the
isovector $B(E2)$.  If the latter is negligible, then indeed one
basically has the same relation between scissors mode excitations and
nuclear collectivity, as empirically observed in the Sm isotopes.

However, the derivation of the above energy weighted sum rule is quite
general and should therefore hold (in the mathematical sense) in all
regions, not just where the deformation is strong.  To best illustrate
the need for the isovector $B(E2)$, consider a nucleus with a close
shell of neutrons or protons.  In such a nucleus, and neglecting
ground state correlations, the scissors mode excitation rate will
vanish as one needs both open shell neutrons and protons to get a
finite scissors mode excitation rate.  On the other hand, the $B(E2,
0_{1} \rightarrow 2_{1})$ can be quite large.  However, if we have say
an open shell of protons and a closed shell of neutrons, the $B(E2,
0_{1} \rightarrow 2_{1})$ can be quite substantial.  Many vibrational
nuclei are of such an ilk, and they have large $B(E2)$'s from ground,
e.g. 20 W.u.

However, in the above circumstances (closed neutron shell), the
neutrons will not contribute 
to the $B(E2)$ even if we give them an effective charge.  But if only
the protons contribute, it is clear that $B(E2, {\rm isovector}) = B(E2,
{\rm isoscalar})$.

As an example, let us consider the even-even Be isotopes $^{6}$Be,
$^{8}$Be, $^{10}$Be and $^{12}$Be.  In so doing, we go far away from
the valley of stability, but this is in tune with modern interests in
radioactive beams.

Fayache, Sharma and Zamick~\cite{Fayache-MS:1996annp} have previously
considered $^{8}$Be 
and $^{10}$Be.  The point was made that these two nuclei had about the
same calculated $B(E2, 0_{1} \rightarrow 2_{1})$, but the isovector
orbital $B(M1)$'s were significantly smaller in $^{10}$Be than in
$^{8}$Be.  This was against the systematic that $B(M1)_{{\rm orbital}}$ is
proportional merely to $B(E2)$.  In detail, the calculated $B(M1, 0_{1}
\rightarrow 1_{1})$ was $(2/\pi) \mu_{N}^{2}$ for $^{8}$Be and in
$^{10}$Be was $(9/32\pi) \mu_{N}^{2}$ ($T=1 \rightarrow T=1)$ and 
$(15/32 \pi) \mu_{N}^{2}$ ($T=1 \rightarrow T=2$).  Thus the ratio of
isovector orbital $B(M1)$'s $^{10}$Be/$^{8}$Be = 3/8.  

The energy
weighted sum rule has been verified in the $0p$ shell by Fayache,
Sharma and Zamick~\cite{Fayache-MS:1996annp}.  Using values 
$\chi = 0.5762$ MeV/fm$^{4}$ 
for $^{8}$Be and 0.3615 MeV/fm$^{4}$ for $^{10}$Be there is agreement
between the left hand side and right hand side of
equation~\ref{eq:ewsr}.  The values are 6.411 MeV for $^{8}$Be and
2.030 MeV for $^{10}$Be.

We now extend the calculations to include $^{6}$Be and $^{12}$Be.
These are singly closed shell nuclei.  We see in table~\ref{tab:BE2}
how everything hangs together.  We can explain the reduction is $B(M1)$
in $^{10}$Be relative to $^{8}$Be by the fact that the isovector
$B(E2)$ in $^{10}$Be is much larger than in $^{8}$Be.  Note that the
isoscalar $B(E2)$'s are almost the same in these two nuclei.  The summed
$B(M1)$ in $^{8}$Be is $2/\pi$, but in $^{10}$Be it is only 3/8 of that.

In $^{6}$Be and $^{12}$Be, the E2 transition is from two protons with
L=0, S=0 to two protons with L=2, S=0.  Note that, surprisingly,. the
coefficients in front of the effective charge factors is larger for
singly-magic $^{6}$Be than it is for the open shell nucleus $^{8}$Be.
The factors are respectively $12.5 b^{4}/\pi$ and $8.75 b^{4}/\pi$.
However, the charge factor for $^{6}$Be ($^{12}$Be) is $e_{p}^{2}$,
whereas for $^{8}$Be it is $(e_{p}+e_{n})^{2}$.  The latter gives a
factor of four enhancement for the isoscalar $B(E2)$ in $^{8}$Be.

Again we see from table~\ref{tab:BE2} that the isoscalar and isovector
B(E2)'s are necessarily the same for $^{6}$Be and $^{12}$Be and, when
this is fed into the sum 
rule of Zheng and Zamick~\cite{Zamick-L:1992prc}, one gets the
consistent result that $B(M1)_{{\rm orbital}}$ is zero for these nuclei.

At about the same time as the work of~\cite{Zamick-L:1991prc} was
performed, the same problem was addressed in the context of IBA-2 by
Heyde and deCoster~\cite{Heyde-K:1991prc}.  More recently, they have
extended the sum rules to include E(0) and M(3)
excitations~\cite{Heyde-K:1994prc}.  Their energy weighted sum
rule~\cite{Heyde-K:1994prc} appears in a somewhat different form than
the one in Ref.~\cite{Zamick-L:1992prc}.
\begin{equation}
\sum_{i} B(M1, 0_{1}^{+} \rightarrow 1_{i}) (E_{i}-E_{0})
  = c \sum_{i} B(E2, 0_{1}^{+} \rightarrow 2_{i}^{+})
\end{equation}
In the above $B(E2)$ is for a purely {\em isoscalar} operator with an
effective charge $T(E2) = e_{{\rm eff}} (Q_{\pi}+Q_{\nu})$.
However their effective charge is proportional to 
$(N_{\pi} N_{\nu}/N^{2})^{1/2}$ where $N_{\pi}$ and $N_{\nu}$ are the
numbers of neutron bosons and proton bosons.  This expression will
therefore also vanish for a single closed shell, for in that case
either $N_{\pi}$ or $N_{\nu}$ will vanish.

\section{Comparison of the Energy Weighted Sum Rule in the
fermion SU(3) and boson SU(3) models}

We now compare the results of the above {\em fermion} SU(3) model with
those of a {\em boson} SU(3) model.
It is easy to show the following relation
\begin{eqnarray}
4 \langle Q_{\pi} \cdot Q_{\nu} \rangle
  & = & \sum_{i} B(E2, e_{p}=1, e_{n}=1) \nonumber \\
  && - \sum_{i} B(E2, e_{p}=1, e_{n}=-1)
\end{eqnarray}
Thus the right hand side of Eq.~(\ref{eq:ewsr}) is proportional to 
$\langle Q_{\pi} \cdot Q_{\nu} \rangle$.  Using the techniques
developed in the SU(3) boson
model~\cite{Rosensteel-G:1990prc}, one can show that
\begin{eqnarray}
Q_{\pi} \cdot Q_{\nu}
  & = & {15 \over 64} \lbrack C_{2}(\lambda, \mu) 
  - C_{2}(\lambda_{\pi}, \mu_{\pi}) - C_{2}(\lambda_{\nu}, \mu_{\nu}) 
    \rbrack \nonumber \\
 &&  - {3 \sqrt{30} \over 32} [C^{(11)}(\pi)\times C^{(11)}(\nu)]_{L=0}^{(22)}
\end{eqnarray}
where $C^{(11)}$ denotes the eight-dimensional irreducible tensor operator 
formed by the generators of the SU(3) subalgebra of the boson algebra U(6),
$C_{2}(\lambda, \mu)
= {2 \over 3} (\lambda^{2} + \mu^{2} + \lambda \mu + 3 \lambda + 3 \mu)$ is 
the SU(3) quadratic Casimir operator, and the coupled operator is a $(22)$
SU(3) coupled tensor operator.
When a $Q \cdot Q$ interaction is present between proton bosons and
neutron bosons the ground state in the SU(3) limit of IBA-2 is
\begin{equation}
|g \rangle = 
  | [N_{\pi}](2N_{\pi},0) [N_{\nu}](2N_{\nu},0);
    (2N_{\pi}+2N_{\nu},0)_{L=0} \rangle
\end{equation}

Evaluating the matrix elements of the $(22)$ tensor using the
SU(3) 9-$(\lambda,\mu)$ recoupling coefficients, 
we obtain the following result
\begin{equation}
\langle g || Q_{\pi} \cdot Q_{\nu} || g \rangle
  = N_{\pi} N_{\nu} \left\lbrack 
    2 + {3 \over 2(N_{\pi}+N_{\nu})-1} \right\rbrack .
\end{equation}

Note that this expression vanishes unless {\em both} $N_{\pi}$ and
$N_{\nu}$ are non-zero., i.e. unless there are both proton bosons and
neutron bosons present.

If we normalize $^{8}$Be to unity, we can compare the fermion SU(3)
model and the boson SU(3) model predictions for $\langle Q_{\pi}
\cdot Q_{\nu} \rangle$.  This is done in table~\ref{tab:QQ}.  This is
equivalent to a comparison of the energy weight sum rule for the
orbital $B(M1)$ strength using the same value of $\chi$ for all the Be
isotopes.

As mentioned before both models correctly predict that for the singly
magic nuclei the expectation value of $\langle Q_{\pi} \cdot Q_{\nu} \rangle$
is zero.  There is however a substantial difference -- more than a
factor of two in the ratio of $^{10}$Be to $^{8}$Be for 
$\langle Q_{\pi} \cdot Q_{\nu} \rangle$.

In a heavier nucleus we might think that such a difference could be
due in part to the presence of $g$ bosons.  However, in the $0p$
shell, we cannot couple two nucleons to $L=4$.  The most plausible
explanation for the discrepancy is that a $Q \cdot Q$ 
interaction between fermions does not imply a $Q \cdot Q$ interaction
between proton bosons and neutron bosons.  For example, in the fermion
case there is a $Q \cdot Q$ interaction between identical particles.

\acknowledgements

This work was supported by a Department of Energy grant
DE-FG02-95ER40940 and by a Stockton College summer R@PD research
grant.

%\newpage
\mediumtext
%\widetext

\begin{table}
\caption{The values of $B(M1)_{{\rm orbital}}$, $B(E2)_{{\rm
isoscalar}}$ and $B(E2)_{{\rm isovector}}$ for Be isotopes.}
\label{tab:BE2}
\begin{tabular}{cccccc} \hline
Nucleus & & $b$ (fm)\tablenotemark[1] & $B(M1)_{{\rm orbital}}$ 
        & $B(E2)_{{\rm isoscalar}}$ & $B(E2)_{{\rm isovector}}$ \\
$^{6}$Be  & & 1.553 & 0 & 15.63\tablenotemark[2] & 15.63\tablenotemark[2] \\
$^{8}$Be  & & 1.597 & $2/\pi = 0.637$ & 73.54\tablenotemark[3] & 6.469 \\
$^{10}$Be & $T=1 \rightarrow T=1$ & 1.635 & $9/32\pi = 0.0895$ 
          & 63.24\tablenotemark[4] & 31.19\tablenotemark[5] \\
          & $T=1 \rightarrow T=2$ & & $15/32\pi = 0.149$ & 0 & 3.203 \\
$^{12}$Be & & 1.669 & 0 & 20.85 & 20.85 \\ \hline
\end{tabular}
\tablenotemark[1]{$b^{2} = 41.46/(\hbar \omega)$, 
$\hbar \omega =  45/A^{2/3} - 25/A^{1/3}$.}

\tablenotemark[2]{The analytic expression in $^{6}$Be is 
$B(E2) = {50 \over 4 \pi} b^{4} e_{p}^{2}$.}

\tablenotemark[3]{The analytic expression in $^{8}$Be is 
$B(E2) = {35 \over 4 \pi} b^{4} (e_{p}+e_{n})^{2}$.}

\tablenotemark[4]{$B(E2)_{{\rm isoscalar}} = 0$ to the $2_{1}^{+}$
state, and is equal to $68.24 e^{2}$ fm$^{4}$ to the $2_{2}^{+}$ state.}

\tablenotemark[5]{$B(E2)_{{\rm isovector}} = 31.19 e^{2}$ fm$^{4}$ to
the $2_{1}^{+}$ state, and is equal to zero to the $2_{2}^{+}$ state.}
\end{table}

\begin{table}
\caption{The ratio $\langle Q_{\pi} \cdot Q_{\nu} \rangle_{^{A}{\rm Be}}/
\langle Q_{\pi} \cdot Q_{\nu} \rangle_{^{8}{\rm Be}}$.}
\label{tab:QQ}
\begin{tabular}{ccc} \hline
Nucleus & Fermion SU(3) & Boson SU(3)\tablenotemark[1]
 \\ \hline
$^{6}$Be & 0 & 0 \\
$^{8}$Be & 1 & 1 \\
$^{10}$Be & ${33 \over 72} = 0.458$ & 1 \\
$^{12}$Be & 0 & 0 \\ \hline
\end{tabular}
\tablenotemark[1]{For $^{8}$Be we have
$N_{\pi} =2, N_{\nu} = 2$; for $^{10}$Be we have the same value
because having four neutrons in the $0p$ shell is equivalent to having
two neutron holes.}
\end{table}

\end{document}